\documentclass[12pt,twoside]{article}
\usepackage{latexsym}
\usepackage{longtable}
\usepackage{epsfig}
\usepackage{a4wide}
\usepackage{amsmath,amsthm,amsfonts,amssymb,bbm}
\usepackage{graphicx,psfrag}
\usepackage{cite}
\numberwithin{equation}{section} 
\usepackage{epstopdf}
\graphicspath{{images/}}

\begin{document}
\begin{titlepage}
\begin{center}
\vspace*{2cm} {\Large {\bf The Crossover Regime for the Weakly\medskip\\
Asymmetric Simple Exclusion Process\bigskip\bigskip\\}}
{\large{Tomohiro Sasamoto$^\star$ and
Herbert Spohn$^\dag$}}\bigskip\bigskip\\
  {$^\star$
  Department of Mathematics and Informatics, 
 Chiba University,\\ 1-33 Yayoi-cho, Inage, Chiba 263-8522, Japan\\
 e-mail:~{\tt sasamoto@math.s.chiba-u.ac.jp\medskip}}

 {$^\dag$  Zentrum Mathematik and Physik Department, TU M\"unchen,\\
 D-85747 Garching, Germany\\e-mail:~{\tt spohn@ma.tum.de}}\\

\end{center}
\vspace{5cm} \textbf{Abstract.} We consider the asymmetric simple
exclusion process in one dimension with weak asymmetry (WASEP) and
0\,-1 step initial condition. Our interest are the fluctuations of
the time-integrated particle current at some prescribed spatial location.
One expects a crossover from Gaussian to Tracy-Widom distributed
fluctuations. The appropriate crossover scale is an asymmetry of
order $\sqrt{\varepsilon}$, times of order $\varepsilon^{-2}$, and a
spatial location of order $\varepsilon^{-3/2}$. For this parameter
window we obtain the limiting distribution function of the
integrated current in terms of an integral over the difference of two
Fredholm determinants. For large times, on the scale
$\varepsilon^{-2}$, this distribution function converges to the one
of Tracy-Widom.
\end{titlepage}

\section{Introduction}\label{in}

The asymmetric simple exclusion process (ASEP) on the
one-dimensional lattice $\mathbb{Z}$ is a stochastic particle system
with at most one particle per site. An ASEP particle waits a unit
exponentially distributed random time and then jumps to the right
with probability $p$, $0\leq p\leq 1$, and to the left with probability
$q= 1-p$. The jump is actually carried out only if the destination
site is empty. Waiting times and jump probabilities are independent.
We will set $p\leq \frac{1}{2}$, so predominantly particles move
towards the left, and always consider 0\,-1 step initial condition
for which the left half lattice is empty and the right half lattice
is completely filled. We label particles from left to right. If
$x_m(t)$ denotes the position of the $m$-th particle,
$m=1,2,\ldots$, then $x_m(0)=m$ and $x_m(t)<x_{m+1}(t)$ for all $t\geq
0$.

In a celebrated work \cite{Jo}, Johansson investigated for the case
$p=0$, the totally asymmetric simple exclusion process (TASEP), the
time-integrated particle current across the origin, denoted here by
$\mathcal{J}(0,t)$. $- \mathcal{J}(0,t)$ is simply the total number of particles which
have jumped across the edge $(0,1)$ up to time $t$. Johansson proves
that in the limit $t\to\infty$ it holds
\begin{equation}\label{1.1}
\mathcal{J}(0,t)\cong -\tfrac{1}{4}t + 2^{-4/3} t^{1/3}\xi_{\mathrm{TW}}\,.
\end{equation}
$\xi_{\mathrm{TW}}$ is a Tracy-Widom distributed random variable,
which first appeared in the context of the large $N$ limit of the
largest eigenvalue of a GUE random matrix \cite{Fo,TW1}. The
asymptotics (\ref{1.1}) has recently been extended by Tracy and
Widom \cite{TW,TW2} to the partially asymmetric simple exclusion
process (PASEP). In this case $\mathcal{J}(0,t)$ is the number of signed jumps
across the edge $(0,1)$ up to time $t$ and, in fact, (\ref{1.1}) still holds with the only
modification that $t$ on the left hand side is replaced by $t/(q-p)$.

In the limit of symmetric jumps, $p=q=\frac{1}{2}$, one finds
\begin{equation}\label{1.2}
\mathcal{J}(0,t)\cong (2\pi)^{-1/4} (\sqrt{2}-1)^{1/2} t^{1/4}\xi_{\mathrm{G}}
\end{equation}
for large $t$, where $\xi_\mathrm{G}$ is Gaussian
distributed with mean 0 and variance 1. Even finer details are established and we refer to
the recent study  \cite{DG} on the large deviations for $\mathcal{J}(0,t)$.

With this perspective it is of interest to understand in more detail
the crossover between the Gaussian central limit theorem
(\ref{1.2}) and the (non-Gaussian) Tracy-Widom statistics
(\ref{1.1}). The appropriate crossover scale was already identified
by Bertini and Giacomin \cite{BG}. For such a study it is convenient
to introduce the dimensionless scale parameter $\varepsilon$,
$\varepsilon >0$ and $\varepsilon\ll 1$, and to consider the time
scale $\varepsilon^{-2} t$, $t=\mathcal{O}(1)$. The strength of
the asymmetry is encoded by the choice
\begin{equation}\label{1.3}
p=\tfrac{1}{2}(1-\beta \varepsilon^\alpha)\,,\quad
\beta>0\,,\alpha>0\,,
\end{equation}
which corresponds to a weakly asymmetric simple exclusion process
(WASEP). The standard WASEP is the particular case $\alpha=1$. The
macroscopic density profile, $\rho(x,t)$, is then governed by the
dissipative Burgers equation
\begin{equation}\label{1.3a}
\frac{\partial}{\partial t}\rho +\frac{\partial}{\partial x}
\Big(-\beta\rho(1-\rho)-\frac{1}{2}\frac{\partial}{\partial x}\rho\Big)=0\,.
\end{equation}
In \cite{DPS,DG1} the associated Gaussian fluctuation theory has
been developed which, in particular, proves that
\begin{equation}\label{1.4}
\mathcal{J}^\varepsilon (0,\varepsilon^{-2} t)=-\frac{1}{4}\beta t
\varepsilon^{-1} + c(t) \varepsilon^{-1/2} \xi_\mathrm{G}
\end{equation}
for small $\varepsilon$. The superscript $\varepsilon$ for $\mathcal{J}^\varepsilon$
should remind that the jump probability is adjusted according to
(\ref{1.3}). The variance $c(t)^2$ can be computed in principle from
fluctuating hydrodynamics \cite{Sp91} which arrives at an expression  in terms of
the linearization of (\ref{1.3a}) around the time evolved step profile.

Following \cite{BG} the crossover scale is $\alpha=\frac{1}{2}$.
This crossover scale has also been noted in the 
spectral gap \cite{DK} of the WASEP generator and in the large deviations of the 
total current \cite{DM,PM}.
We expect that, with the appropriate adjustment of constants,
(\ref{1.4}) holds whenever $\alpha >\frac{1}{2}$. On the other side
for $\alpha<\frac{1}{2}$ the limiting distribution should be
Tracy-Widom. In this paper we investigate only the crossover regime
$\alpha=1/2$, for which an added interest comes from the relation to
the KPZ equation, see  \cite{BG,BQS} and the discussion in our companion
papers \cite{SS,SS1}. Hence we fix
\begin{equation}\label{1.5}
p=\frac{1}{2}\big(1-\beta\sqrt{\varepsilon}\big)\,,\quad
q-p=\beta\sqrt{\varepsilon}\,,\quad \tau=\frac{p}{q}\cong
1-2\beta\sqrt{\varepsilon}\,.
\end{equation}
On the time scale $\varepsilon^{-2}t$ the average time integrated
current is of order $\varepsilon^{-3/2}$ and a typical particle
profile increases linearly over the interval $[-\beta
t\varepsilon^{-3/2}, \beta t\varepsilon^{-3/2}]$ with 0 to the left
and 1 to the right of this interval. The one-point distribution of the
time-integrated current will be studied at a general location and
not only at the origin.

In Section \ref{sec5} we summarize our main result, which states that in the rescaled units the fluctuations of the integrated current have a size of order $t^{1/3}$ with a $t$-dependent amplitude $\xi_t$ of order 1.
The distribution function of $\xi_t$ is given in (\ref{4.25}), from which it is easily checked that $\xi_t$
is Tracy-Widom distributed  in the limit $t\to\infty$. Thus at the crossover scale, one still has the same long time behavior as for the PASEP.
 
 The analysis heavily relies on the methods
developed by Tracy and Widom in \cite{TW}. In Section \ref{sec2} we employ
the Ramanujan summation formula, an observation which will be
instrumental in the asymptotic analysis. The saddle point is
discussed in Section \ref{sec3}, while in Section \ref{sec4} we study the
$\mu$-integration and convert the contour integrations to a Fredholm
determinant in $L^2(\mathbb{R})$ with a real symmetric kernel. In Appendix \ref{C}
we argue that one could also perform the $\mu$-integration in the very first step, still to arrive at the same result.


\section{The Tracy and Widom contour integration formula for the
particle's positions}\label{sec2}
\setcounter{equation}{0}

As discussed in \cite{TW2}, the time-integrated current is directly
linked to the motion of an ASEP particle with its label properly
adjusted. Hence, our focus will be on the motion of
particles. To ease the comparison we follow closely the notation in \cite{TW},
which will be referred  to merely as TW. We introduce the dimensionless parameter $\sigma$,
$0<\sigma<1$, to label the reference point. At time
$\varepsilon^{-2}t$ the particle index of interest equals
\begin{equation}\label{1.6}
    m=\sigma\beta t \varepsilon^{-3/2}
\end{equation}
and the $m$-th particle is typically at the location
\begin{equation}\label{1.7}
c_1 \varepsilon^{-3/2}\,,\quad c_1=(-1+2\sqrt{\sigma})\beta t\,.
\end{equation}
In fact we will have to include a subleading correction to $c_1
\varepsilon^{-3/2}$ of order $\varepsilon^{-1/2} \log \varepsilon$.
The fluctuation scale is
\begin{equation}\label{1.8}
c_2 \varepsilon^{-1/2}\,,\quad
c_2=\sigma^{-1/6}(1-\sqrt{\sigma})^{2/3}(\beta t)^{1/3}\,.
\end{equation}
For later use we also introduce 
\begin{eqnarray}\label{1.9}
&&c_3=\sigma^{-1/6}(1-\sqrt{\sigma})^{5/3}(\beta t)^{1/3}\,,\quad
c_4=\sqrt{\sigma}/(1-\sqrt{\sigma})\,, \nonumber\\[1ex]
&&\gamma_t=2\beta c_3 c_4=2\beta(\beta
t)^{1/3}(\sqrt{\sigma}(1-\sqrt{\sigma}))^{2/3}\,.
\end{eqnarray}

With these conventions our task will be to study the limit
$\varepsilon\to 0$ of the WASEP distribution function
\begin{equation}\label{1.10}
F^\varepsilon_t(s) =\mathbb{P}\big(x_m(\varepsilon^{-2}t)-c_1\varepsilon^{-3/2}-(c_2/\gamma_t)\varepsilon^{-1/2}
\log(2\beta\sqrt{\varepsilon})\leq c_2 s \varepsilon^{-1/2}\big)\,.
\end{equation}

TW start their analysis with
the identity
\begin{equation}\label{2.1}
\mathbb{P}\big(x_m(t/(q-p))\leq
x\big)=\int_{\mathcal{C}_0}\prod^\infty_{k=0}(1-\mu\tau^k)\det
(1+J(\mu))\frac{\mathrm{d}\mu}{\mu}\,,
\end{equation}
see TW (25), (27), and Lemma 4. We follow their convention that all contour integrals are given a factor
$1/2\pi\mathrm{i}$. 
$\mathcal{C}_0$ is a circle with center at 0 and radius in the open
interval $(\tau,1)$. The operator $J(\mu)$ has the kernel
$J(\mu;\eta,\eta')$ given by
\begin{equation}\label{2.2}
J(\mu;\eta,\eta')=\int_{\mathcal{C}_1}\frac{\varphi_\infty(\zeta)}{\varphi_\infty(\eta')}
\frac{\zeta^m}{(\eta')^{m+1}}\frac{\mu f(\mu,\zeta/\eta')}{\zeta-\eta}\mathrm{d} \zeta
\end{equation}
with $\mu\in\mathbb{C}$. Here $\eta,\eta'$ are on a circle with center 0 and radius
$r\in(\tau,1)$ and, as a linear operator, $J(\mu)$  acts on functions
on this circle. The integration contour $\mathcal{C}_1$ is over a
circle with center 0 and radius in the interval $(1,r/\tau)$.
$\varphi_\infty$ is defined through
\begin{equation}\label{2.3}
\varphi_\infty(\eta)=(1-\eta)^{-x} \mathrm{e}^{t(\eta/(1-\eta))}
\end{equation}
and $f$ through
\begin{equation}\label{2.4}
f(\mu,z)=\sum^\infty_{k=-\infty} \frac{\tau^k}{1-\mu\tau^k}z^k\,.
\end{equation}

By immediate bounds, for $\mu \in \mathbb{C}\backslash\{0,\tau^n, n\in\mathbb{Z}\}$ the function $f(\mu,z)$ is analytic in the annulus
$1<|z|<\tau^{-1}$. We will need its analytic extension, which can be
deduced from a product formula representation. Following \cite{A},
we set
\begin{equation}\label{2.5}
(a;q)_\infty=\prod^\infty_{n=0}(1-a q^n)\,,\quad |q|<1\,,\quad (a;q)_n=(a;q)_\infty/(aq^n;q)_\infty\,.
\end{equation}
The Ramanujan summation formula, see \cite{A}, Theorem 10.5.1, states
\begin{equation}\label{2.6}
\sum^\infty_{n=-\infty} \frac{(a;q)_n}{(b;q)_n}x^n
=\frac{(ax;q)_\infty (q/ax;q)_\infty (q;q)_\infty (b/a;q)_\infty}
{(x;q)_\infty (b/ax;q)_\infty (b;q)_\infty (q/a;q)_\infty}\,,
\end{equation}
provided $|q| < 1$ and $|b/a| < |x| < 1$. Setting $a=\mu$, $b=\mu \tau$, $x=\tau z$, $q=\tau$, one easily checks that
for $1 < |z| < \tau^{-1}$ it holds
\begin{eqnarray}\label{2.7} 
&&\hspace{-46pt} \mu f(\mu,z)=\mu \frac{(\mu\tau z;\tau)_\infty
(1/\mu z;\tau)_\infty (\tau;\tau)_\infty (\tau;\tau)_\infty} {(\tau
z ;\tau)_\infty (1/z;\tau)_\infty (\mu;\tau)_\infty
(\tau/\mu;\tau)_\infty}\nonumber\\
&&\hspace{-36pt}=\frac{1-\mu z}{(1-z)(1-\mu)}\prod^\infty_{n=1}
\frac{(1-\tau^n)(1-\tau^n)}{(1-z\tau^n)(1-z^{-1}\tau^n)}
\prod^\infty_{n=1} \frac{(1-\mu z\tau^n)(1-(\mu
z)^{-1}\tau^n)}{(1-\mu\tau^n)(1-\mu^{-1}\tau^n)} \,.
\end{eqnarray}
Since $0 < \tau < 1$, the right hand side is analytic in $z$ and $\mu$ in the domain $\mathbb{C}\backslash\{0,\tau^n, n\in\mathbb{Z}\}$. Hence the right hand side of (\ref{2.7}) is the analytic continuation of $f$ as defined through (\ref{2.4}).  In both variables
$f$ has
simple poles at $\tau^n$, 
$n\in\mathbb{Z}$.


\section{Saddle point analysis and limit kernel}\label{sec3}
 \setcounter{equation}{0}

We investigate the limit of the kernel (\ref{2.2}) of $J(\mu)$ as
$\varepsilon\to 0$. The integrand is written as the product of
three factors,
\begin{equation}\label{3.1}
 \frac{\varphi_\infty
 (\zeta)\zeta^m}{\varphi_\infty(\eta')(\eta')^m} \times\frac{1}{\eta'(\zeta-\eta)} \times\mu
 f(\mu,\zeta/\eta')=Q_1\times Q_2\times Q_3\,.
 \end{equation}
We study each factor separately and start with 
$Q_1$.

In (\ref{1.10}) the logarithmic term is switched to the right as
\begin{equation}\label{3.2a}
\mathbb{P}\Big(x_m(\varepsilon^{-2} t) -c_1 \varepsilon^{-3/2}\leq
c_2\big(s+\gamma_t^{-1} \log
(2\beta\sqrt{\varepsilon})\big)\varepsilon^{-1/2}\Big)
\end{equation}
and thus regarded as a shift of $s$. First we will ignore this
shift, which can be included later on because of the uniform error
estimates. Hence the parameters in the numerator $\varphi_\infty(\zeta)\zeta^m$
of $Q_1$ are
\begin{equation}\label{3.3}
m=\sigma \beta t \varepsilon^{-3/2}\,,\quad x=c_1 \varepsilon^{-3/2}
+ c_2 s \varepsilon^{-1/2}\,.
\end{equation}
As a consequence, the saddle point analysis is identical to the one
of TW with $t$ replaced by $\varepsilon^{-3/2}$. The saddle point is given by
\begin{equation}\label{3.4}
\xi=-c_4\,,
\end{equation}
see (\ref{1.9}) for the definition. Setting
\begin{equation}\label{3.5}
\varphi_\infty(\zeta)\zeta^m=\varphi_\infty(\xi)\xi^m \mathrm{e}^{\psi(\zeta)}\,,
\end{equation}
in a neighborhood of $\zeta=\xi$ one finds, see TW (30),
\begin{eqnarray}\label{3.6}
&&\hspace{-66pt}\psi(\zeta)=-(c_3)^3\varepsilon^{-3/2}(\zeta-\xi)^3/3 +c_3 s
\varepsilon^{-1/2}(\zeta-\xi)\nonumber\\[1ex]
&&\hspace{-26pt}+\mathcal{O}(\varepsilon^{-3/2}(\zeta-\xi)^4) +
\mathcal{O}(\varepsilon^{-1/2}(\zeta-\xi)^2)\,.
\end{eqnarray}

The rescaling close to the saddle point corresponds to the
substitutions
\begin{equation}\label{3.7}
\eta\to \xi +c^{-1}_3 \sqrt{\varepsilon}\eta\,, \quad
\eta'\to \xi +c^{-1}_3 \sqrt{\varepsilon}\eta'\,,\quad
\zeta\to\xi + c^{-1}_3 \sqrt{\varepsilon} \zeta\,.
\end{equation}
Then
\begin{equation}\label{3.8}
\lim_{\varepsilon\to 0} \frac{\varphi_\infty
 (\zeta)\zeta^m}{\varphi_\infty(\eta')\eta'^m}=\exp[-\tfrac{1}{3} \zeta^3 +
 \tfrac{1}{3} (\eta')^3 + s(\zeta-\eta')]\,.
\end{equation}
For the limit in (\ref{3.8}), $\zeta\in \Gamma_\zeta$ and $\eta'\in
\Gamma_\eta$, where $\Gamma_\zeta$ consists of the two rays from $-c_3$
to $-c_3 +\infty \mathrm{e}^{\pm 2\pi i/3}$, while $\Gamma_\eta$ consists of
the two rays from 0 to $+\infty \mathrm{e}^{\pm \pi i/3}$.

Close to the saddle point the second factor of (\ref{3.1}) reads
\begin{equation}\label{3.8a}
Q_2= - \frac{c_3}{c_4\sqrt{\varepsilon}(\zeta-\eta)}\,.
\end{equation}

The factor $Q_3$ needs more work. By (\ref{3.7}) the ratio $\zeta/\eta'$ close to the saddle point reads 
\begin{equation}\label{3.9}
\frac{\xi+c^{-1}_3 \zeta\sqrt{\varepsilon}}{\xi+c^{-1}_3
\eta'\sqrt{\varepsilon}}=1 +(c_3 c_4)^{-1} (\eta'-\zeta)
\sqrt{\varepsilon}+ \mathcal{O}(\varepsilon) \,.
\end{equation}
Hence, according to (\ref{3.1}),  $\mu f(\mu, \cdot)$ has to be evaluated at $1+\sqrt{\varepsilon}z$
with
\begin{equation}\label{3.9b}
z = (c_3 c_4)^{-1} (\eta'-\zeta)\,.
\end{equation}
Correspondingly  in  (\ref{2.7}) we 
substitute
$z$ by $1+\sqrt{\varepsilon}z$ and write the product as $Q_4\times Q_5\times Q_6$.
Then
\begin{equation}\label{3.9a}
\mu
f(\mu, 1+\sqrt{\varepsilon}z) = Q_4Q_5Q_6\,.
\end{equation}
We study the limit of each factor as $\varepsilon \to 0$.

The factor $Q_4$ reads
\begin{equation}\label{3.10}
Q_4=\frac{1-\mu(1+\sqrt{\varepsilon}z)}{-\sqrt{\varepsilon}z
(1-\mu)}=\frac{1}{-\sqrt{\varepsilon}z}
(1+\mathcal{O}(\sqrt{\varepsilon}))\,.
\end{equation}

To study the limit of the factor $Q_5$ we introduce the
$q$-gamma function,
\begin{equation}\label{3.11}
\Gamma_q (x)= \frac{(q;q)_\infty}{(q^x;q)_\infty}(1-q)^{1-x}\,,\quad
\textrm{when } |q|<1\,,
\end{equation}
see \cite{A} (10.3.3). Setting $q^x =1+\sqrt{\varepsilon}z$ one arrives at the identity
\begin{eqnarray}\label{3.12}
 &&\hspace{-16pt}   Q_5= \prod^\infty_{n=1} \frac{(1-\tau^n)(1-\tau^n)}
    {(1- (1+\sqrt{\varepsilon}z)\tau^n)(1-(1+\sqrt{\varepsilon}z)^{-1}\tau^n)}
    \nonumber\\[1ex]
&&\hspace{2pt} = -\Gamma_\tau(x) \Gamma_\tau
    (-x)(1+\sqrt{\varepsilon}z)^{-1}(\sqrt{\varepsilon}z)^2(1-\tau)^{-2}\,.
\end{eqnarray}
Since $q=\tau=1-2\beta\sqrt{\varepsilon}$, it follows that
$x=-(z/2\beta)+\mathcal{O}(\sqrt{\varepsilon})$. The $q$-gamma
function converges to the gamma function, $\Gamma$, in the
limit $q\to 1$. Hence the limit $\varepsilon\to 0$ on the right hand side of  (\ref{3.12})
becomes
\begin{equation}\label{3.13}
-\Gamma(z/2\beta) \Gamma(-z/2\beta) (z/2\beta)^2\,.
\end{equation}
Using
\begin{equation}\label{3.14}
-z \Gamma(-z)= \Gamma(-z+1)\,,\quad \Gamma(z) \Gamma (1-z)=\frac{\pi
}{\sin (\pi z)}\,,
\end{equation}
yields
\begin{equation}\label{3.15}
\lim_{\varepsilon\to 0}\prod^\infty_{n=1}
\frac{(1-\tau^n)(1-\tau^n)}{(1-(1+\sqrt{\varepsilon}z)\tau^n)(1-(1+\sqrt{\varepsilon}z)^{-1}\tau^n)}= \frac{\pi
z/2\beta}{\sin(\pi z/2\beta)}\,.
\end{equation}

The factor $Q_6$ reads
\begin{eqnarray}\label{3.16}
&&\hspace{-25pt}Q_6= \exp\Big[\sum^\infty_{n=1}\Big(\log\frac{1-\mu (1+\sqrt{\varepsilon}z)
\tau^n}{1-\mu\tau^n}+\log\frac{1-(\mu (1+\sqrt{\varepsilon}z))^{-1}
\tau^n}{1-\mu^{-1}\tau^n}\Big)\Big]\nonumber\\
&&\hspace{-8pt}=\exp\Big[\sum^\infty_{n=1}\Big(\log\big(1-\sqrt{\varepsilon}z
\frac{\mu\tau^n}{1-\mu\tau^n}\big)+\log\big(1+\frac{\sqrt{\varepsilon}z}{1 + \sqrt{\varepsilon}z}
\frac{\tau^n}{\mu-\tau^n}\big)\Big)\Big]\,.
\end{eqnarray}
Since $\mu\in\mathbb{C}\backslash\mathbb{R}_+$, the argument of the
log lies in $\mathbb{C}\backslash\mathbb{R}_-$. Hence log is
understood as the main branch of the logarithm on
$\mathbb{C}\backslash\mathbb{R}_-$.

\begin{figure}[t]
\begin{center}
\includegraphics[scale=1.]{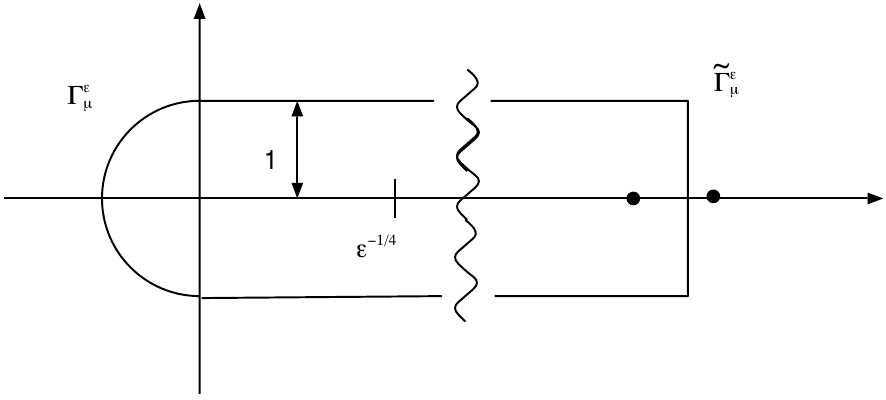}
\end{center}
\caption{The contour
$\Gamma^\varepsilon_\mu \cup\tilde{\Gamma}^{\varepsilon}_\mu$. The dots are the poles at 
$(2\beta\sqrt{\varepsilon})^{-1}\tau$ and $(2\beta\sqrt{\varepsilon})^{-1}$.} 
\end{figure}

Let us define the domain
\begin{equation}\label{3.17}
\mathcal{D}_\kappa =\Big\{\mu\in\mathbb{C}\big| \sup_{0\leq y\leq
1}\big|\frac{\mu y}{1-\mu y}\big|\leq 1\,,\; \sup_{0\leq y\leq
1}|\frac{y}{\mu-y}\big|\leq 1+ \kappa\Big\}
\end{equation}
with $\kappa>0$. If in (\ref{3.16}) we assume that
$\mu\in\mathcal{D}_\kappa$, then the logarithm can be
expanded  since $0<\tau<1$. Hence the
exponent $[\cdot]$ in (\ref{3.16}) reads
\begin{equation}\label{3.18}
\frac{z}{2\beta}\Big(-\int^1_0 dy \frac{\mu}{1-\mu
y} + \int^1_0 dy \frac{1}{\mu- y}\Big)
+R(\varepsilon)= \frac{z}{2\beta} \log (-\mu)
+R(\varepsilon)\,.
\end{equation}
The error term $R(\varepsilon)$ is bounded by
\begin{equation}\label{3.18a}
|R(\varepsilon)| \leq c \sqrt{\varepsilon}(1/2\beta)|z|^2\Big(-\int^1_0 dy \frac{\mu y}{(1-\mu
y)^2} + \int^1_0 dy \frac{y}{(\mu- y)^2}\Big)
\end{equation}
with some constant $c$ independent of $\varepsilon$. According to the definition (\ref{3.9a}), we conclude that, uniformly on $\mathcal{D}_\kappa$ with an error of
order $\sqrt{\varepsilon}\log\varepsilon$,
\begin{equation}\label{3.19}
\lim_{\varepsilon\to 0} -\sqrt{\varepsilon}\mu
f(\mu,1+\sqrt{\varepsilon}z)=\frac{\pi/2\beta}{\sin (\pi z/2\beta)}
\mathrm{e}^{(z/2\beta)\log(-\mu)}\,.
\end{equation}

Since $\tau=1-2\beta\sqrt{\varepsilon}$, the product in (\ref{2.1})
converges to a non-degenerate limit only if
$\mu=\mathcal{O}(\sqrt{\varepsilon})$. Therefore, 
as the final step, we substitute $\mu$ by $2\beta\mu\sqrt{\varepsilon}$.
Then the $\mu$-integration is over a circle with center 0 and radius
in the interval $(2\beta\sqrt{\varepsilon})^{-1}(\tau,1)$. This contour is
deformed to $\Gamma^\varepsilon_\mu \cup\tilde{\Gamma}^{\varepsilon}_\mu$.
$\Gamma^\varepsilon_\mu=\{\mu|\mathrm{dist}
(\mu,\mathbb{R}_+)=1\}\cap\{\Re \mu \leq \varepsilon^{-1/4}\}$ and $\tilde{\Gamma}^{\varepsilon}_\mu$
closes the contour, see Fig. 1, where in brackets we remark that
$\varepsilon^{-1/4}$ and 1 are taken here only for concreteness. As explained in Appendix A, one can
choose $\varepsilon_0>0$ such that, for all
$0<\varepsilon<\varepsilon_0$,
$\Gamma^\varepsilon_\mu\cap\{\mu|\Re
\mu<\kappa\}\subset\mathcal{D}_\kappa$. Also
$\Gamma^\varepsilon_\mu$ converges as $\varepsilon\to 0$ to
$\Gamma_\mu=\{\mu|\mathrm{dist} (\mu,\mathbb{R}_+)=1\}$. By
uniformity in $\varepsilon$ one concludes that for $\mu\in\Gamma_\mu\cap \{\mu|\Re
\mu<\kappa\}$ it holds
\begin{eqnarray}\label{3.20}
&&\hspace{-56pt} 2\beta\mu\sqrt{\varepsilon}
f(2\beta\mu\sqrt{\varepsilon},
1+\sqrt{\varepsilon}z)=-\frac{1}{\sqrt{\varepsilon}}\frac{\pi/2\beta}{\sin(\pi
z/2\beta)}\nonumber\\
&&\hspace{26pt} \times \exp\big[\frac{z}{2\beta}\big(\log(-\mu)+\log
(2\beta\sqrt{\varepsilon})\big)\big]\big(1+\mathcal{O}(\sqrt{\varepsilon})\big)
\end{eqnarray}
with $z$ defined in (\ref{3.9b}).

To complete the argument, one introduces the logarithmic shift of
$s$ and sets
\begin{equation}\label{3.21}
m=\sigma\beta t \varepsilon^{-3/2}\,,\;
x=c_1\varepsilon^{-3/2}+c_2(s+\gamma_t^{-1}
\log(2\beta\sqrt{\varepsilon}))\varepsilon^{-1/2}\,.
\end{equation}
The rescaled kernel of $J(\mu)$ reads then
\begin{equation}\label{3.22}
J^\varepsilon(\mu;\eta,\eta') = J(2\beta\mu\sqrt{\varepsilon};\xi +c^{-1}_3
\eta\sqrt{\varepsilon}, \xi+c^{-1}_3 \eta'\sqrt{\varepsilon})
c^{-1}_3\sqrt{\varepsilon}\,.
\end{equation}
Combining the saddle point asymptotics (\ref{3.8}) with (\ref{3.20}),
it holds pointwise
\begin{equation}\label{3.23a}
\lim_{\varepsilon\to 0} J^\varepsilon(\mu;\eta,\eta')
=
I(\mu;\eta,\eta')
\end{equation}
for $\mu\in\Gamma^\varepsilon_\mu$,
$\Re \mu<\kappa$, where the limit kernel is defined by
\begin{eqnarray}\label{3.24}
&&\hspace{-56pt} I(\mu;\eta,\eta')=\int_{\Gamma_\zeta}
\exp\big[-\tfrac{1}{3} \zeta^3 +\tfrac{1}{3}(\eta')^3 +s
(\zeta-\eta')\big]\frac{1}{\zeta-\eta}\nonumber\\
&&\hspace{26pt} \times
\frac{\pi}{\sin(\gamma_t^{-1}\pi(\eta'-\zeta))}\mathrm{e}^{\gamma_t^{-1}(\eta'-\zeta)\log(-\mu)}\gamma_t^{-1}
\mathrm{d}\zeta
\end{eqnarray}
for $\mu\in\Gamma_\mu$.

As discussed in Appendix B, the operator
$I(\mu)$ is trace class for $\mu\in\Gamma_\mu$. Hence its Fredholm determinant is
well-defined. What we would like to show is the validity of the
limit
\begin{equation}\label{3.26}
\lim_{\varepsilon\to 0} \det (1+J^\varepsilon(\mu))=\det
(1+I(\mu))\,.
\end{equation}
For this we would need the convergence $J^\varepsilon(\mu)\to
I(\mu)$ as $\varepsilon\to 0$ in trace norm. TW have to handle the
same problem for the limiting case $\log
(-\mu)=0$ and the sine expanded as $\gamma_t^{-1}\pi(\eta'-\zeta)$,
which formally corresponds to  $\gamma_t\to\infty$. They use
that $\Re (\eta'-\zeta)>0$, hence their kernels are bounded. In
our case the kernel is singular at $\eta'-\zeta=\gamma_t n$,
$n\in\mathbb{Z}$. This makes the issue of convergence in trace norm
somewhat delicate. 

In the following we assume the validity of the limit in
(\ref{3.26}). We also assume the exponential bounds
\begin{equation}\label{3.27}
|\det (1+J^\varepsilon(\mu))|\leq c \mathrm{e}^{c_0|\mu|}\,,\,|\det
(1+I(\mu))|\leq c \mathrm{e}^{c_0|\mu|}
\end{equation}
for some constant $c_0<1$.


\section{The $\mu$-integration, Fredholm determinant}\label{sec4}
\setcounter{equation}{0}

Rescaling the $\mu$-integration of (\ref{2.1}) as in (\ref{3.22}), one obtains 
\begin{equation}\label{4.0}
F_t^\varepsilon(s) = \int_{(2\beta\sqrt{\varepsilon})^{-1}\mathcal{C}_0}\prod^\infty_{k=0}
\big(1-2\beta\mu\sqrt{\varepsilon}\tau^k\big)\det (1+J^\varepsilon(\mu))\frac{1}{\mu}\mathrm{d}\mu\,.
\end{equation}
For $\mu\in\mathbb{C}\backslash\mathbb{R}_+$ and small $\varepsilon$
it holds
\begin{equation}\label{4.1}
\prod^\infty_{k=0}(1-\mu\tau^k)\cong
\exp\Big[\frac{1}{2\beta\sqrt{\varepsilon}}\int^1_0 \mathrm{d}y
\frac{1}{y}\log(1-\mu y)\Big]\,.
\end{equation}
Thus only if $|\mu|=\mathcal{O}(\sqrt{\varepsilon})$ there is a non-degenerate
limit and \begin{equation}\label{4.2}
\lim_{\varepsilon\to 0}
\prod^\infty_{k=0}\big(1-2\beta\mu\sqrt{\varepsilon}\tau^k\big)=\mathrm{e}^{-\mu}\,.
\end{equation}
Larger values of $|\mu|$  are exponentially
suppressed as $\exp[-1/\sqrt{\varepsilon}\,]$.
We now choose $\kappa$ so large that by assumptions (\ref{3.27}) and by (\ref{4.1})
the error is small uniformly in $\varepsilon$. Then
\begin{equation}\label{4.3}
 \lim_{\varepsilon\to 0} F_t^\varepsilon(s)
=\int_{\Gamma_\mu} \mathrm{e}^{-\mu} \det
(1+I(\mu))\frac{1}{\mu}\mathrm{d}\mu
=F_t(s)\,,
\end{equation}
which defines the limiting distribution function $F_t(s)$ still
depending on the rescaled time parameter $t$.

In principle (\ref{4.3}) is already the final answer, but we still have to
transform to a more manageable form. Let us introduce the kernel
\begin{eqnarray}\label{4.4}
&&\hspace{-30pt}K(\eta_i,\zeta_i;\eta_j,\zeta_j)=
\frac{1}{\zeta_j-\eta_i}\exp\big[-\tfrac{1}{3}\zeta^3_j+\tfrac{1}{3}\eta^3_j
+s(\zeta_j-\eta_j)\big] \frac{\gamma_t^{-1}\pi}{\sin(\gamma_t^{-1}\pi(\eta_j-\zeta_j))}\nonumber\\
&&\hspace{46pt}= \frac{1}{\zeta_j-\eta_i}\exp\big[s(\zeta_j-\eta_j)\big] F(\eta_j,\zeta_j)\,.
\end{eqnarray}
We expand the Fredholm determinant of (\ref{4.3}). The $n$-th
term of the expansion reads
\begin{eqnarray}\label{4.5}
&&\hspace{-30pt}\frac{1}{n!}\int_{\Gamma_\mu} \mathrm{d}\mu
\frac{1}{\mu}\mathrm{e}^{-\mu} \int_{\Gamma_\eta} \mathrm{d}\eta_1\ldots
\mathrm{d}\eta_n \int_{\Gamma_\zeta} \mathrm{d}\zeta_1\ldots \mathrm{d}\zeta_n 
\nonumber\\&&\hspace{16pt}
\prod_{j=1}^{n}\big\{\mu^{\gamma_t^{-1}(\eta_j-\zeta_j)}
\mathrm{e}^{-\mathrm{i}\gamma_t^{-1}\pi(\eta_j-\zeta_j)}\big\}\det
\{K(\eta_i,\zeta_i;\eta_j,\zeta_j)\}_{i,j=1,\ldots,n}\,.
\end{eqnarray}
For this expression the $\mu$-integration can be carried out. We set
\begin{equation}\label{4.7}
w=\gamma_t^{-1} \sum^n_{j=1} (\eta_j-\zeta_j)\,.
\end{equation}
and note that $\Re w >0$, since
for $\zeta_j\in \Gamma_\zeta$, $\eta_j\in \Gamma_\eta$ it holds
$\Re(\eta_j-\zeta_j)>0$.
Therefore 
\begin{eqnarray}\label{4.6}
&&\hspace{-71pt}\mathrm{e}^{-\mathrm{i}\pi w} \int_{\Gamma_\mu} \mathrm{e}^{-\mu} \mu^{w-1}
\mathrm{d}\mu =\frac{1}{2\pi \mathrm{i}} (\mathrm{e}^{\mathrm{i}\pi w}-\mathrm{e}^{-\mathrm{i}\pi w})
\Gamma(w)
\nonumber\\&&\hspace{36pt}
= \frac{1}{2 \pi \mathrm{i}}\int^\infty_0 \mathrm{d}v \frac{1}{v}\mathrm{e}^{-v}
(\mathrm{e}^{\mathrm{i}\pi w}- \mathrm{e}^{-\mathrm{i}\pi w}) v^w\,.
\end{eqnarray}
In addition
\begin{equation}\label{4.9}
\frac{1}{\zeta_j-\eta_i} \mathrm{e}^{s(\zeta_j-\eta_i)}=
-\int^\infty_s \mathrm{d}x \mathrm{e}^{x(\zeta_j-\eta_i)}\,,
\end{equation}
since $\Re(\zeta_j-\eta_i)<0$.

Let us first define the operators
$K^\pm_v$ with integral kernels 
\begin{eqnarray}\label{4.11}
&&\hspace{-46pt}K^\pm_v(x,y)= \int_{\Gamma_\eta} \mathrm{d}\eta
\int_{\Gamma_\zeta} \mathrm{d}\zeta
\frac{\gamma_t^{-1}\pi}{\sin(\gamma_t^{-1}\pi(\eta-\zeta))} \exp \big[-\tfrac{1}{3}
\zeta ^3 +\tfrac{1}{3} \eta ^3\nonumber\\[1ex]
&&\hspace{16pt} + \zeta y -\eta x + \gamma_t^{-1} (\eta-\zeta) \log v
\pm \mathrm{i}\gamma_t^{-1}\pi(\eta-\zeta)\big]\,.
\end{eqnarray}
We expand the determinant in (\ref{4.5}) into cycles. The following rearrangement, illustrated for a 3-cycle, is 
performed for each cycle. We consider the summand $\mathrm{e}^{\mathrm{i}\pi w}v^w$ of (\ref{4.6})
 using the idendity (\ref{4.9}) and the definition (\ref{4.4}) of $F$. Then the term of interest reads
\begin{eqnarray}\label{4.9a}
&&\hspace{-20pt} \int d\eta_1 \mathrm{d}\eta_2\mathrm{d}\eta_3
\int \mathrm{d} \zeta_1 \mathrm{d}\zeta_2 \mathrm{d}\zeta_3
\prod_{j=1}^3\big\{  \mathrm{e}^{\mathrm{i}\pi  \gamma_t^{-1}(\eta_j-\zeta_j)} v^{ \gamma_t^{-1}(\eta_j-\zeta_j)}                              
F(\eta_j,\zeta_j)\big\}
\nonumber\\
&&\hspace{26pt} \times\int^\infty_s \mathrm{d}x_1
\int^\infty_s \mathrm{d}x_2 \int^\infty_s \mathrm{d}x_3 (-1)^3
\mathrm{e}^{x_1(\zeta_2-\eta_1)}\mathrm{e}^{x_2(\zeta_3-\eta_2)}
\mathrm{e}^{x_3(\zeta_1-\eta_3)} \nonumber\\[1ex]
&&\hspace{-6pt} =\int^\infty_s \mathrm{d}x_1 \int^\infty_s \mathrm{d}x_2 \int^\infty_s
\mathrm{d}x_3 (-1)^3\int \mathrm{d}\eta_1 \mathrm{d}\zeta_1  \mathrm{e}^{\mathrm{i}\pi  \gamma_t^{-1}(\eta_1-\zeta_1)} v^{ \gamma_t^{-1}(\eta_1-\zeta_1)} \nonumber\\[1ex]
&&\hspace{26pt}\times F(\eta_1,\zeta_1) \mathrm{e}^{(\zeta_1 x_3
-\eta_1 x_1)}\int \mathrm{d}\eta_2 \mathrm{d}\zeta_2  \mathrm{e}^{\mathrm{i}\pi  \gamma_t^{-1}(\eta_2-\zeta_2)} v^{\gamma_t^{-1} (\eta_2-\zeta_2)} F(\eta_2,\zeta_2)
\mathrm{e}^{(\zeta_2 x_1 -\eta_2 x_2)}
\nonumber\\&&\hspace{26pt}
 \times\int \mathrm{d}\eta_3 \mathrm{d}\zeta_3  \mathrm{e}^{\mathrm{i}\pi  \gamma_t^{-1}(\eta_3-\zeta_3)} v^{\gamma_t^{-1} (\eta_3-\zeta_3)} F( \eta_3,\zeta_3)
\mathrm{e}^{(\zeta_3
x_2 -\eta_3 x_3)}\nonumber\\
&&\hspace{-6pt}=(-1)^3\int^\infty_s \mathrm{d}x_1 \int^\infty_s \mathrm{d}x_2 \int^\infty_s
\mathrm{d}x_3 K_v^+(x_3,x_1) K_v^+(x_1,x_2) K_v^+(x_2,x_3)\,.
\end{eqnarray}
The summand 
$\mathrm{e}^{-\mathrm{i}\pi w}v^w$ of (\ref{4.6}) results correspondingly with $K_v^+$ replaced by $K_v^-$.

We thus expand the Fredholm determinant in (\ref{4.3}), with the $n$-th term of the expansion given in (\ref{4.5}), 
and integrate over $\mu$. According to (\ref{4.6}) this yields the two terms corresponding to $\pm \mathrm{e}^{\pm\mathrm{i}\pi w}v^w$. Using the identity (\ref{4.9a}) and resumming the series results in the difference of two Fredholm determinants. In this difference the constant term 1 cancels. Altogether one therefore obtains
\begin{eqnarray}\label{4.10}
&&\hspace{-32pt}F_t(s)= \int_{\Gamma_\mu}  \mathrm{e}^{-\mu} \det (1+I(\mu))\frac{1}{\mu}\mathrm{d}\mu
\nonumber\\
    &&\hspace{6pt} = 1+\frac{1}{2\pi \mathrm{i}} \int^\infty_0 \mathrm{d}v \frac{1}{v}\mathrm{e}^{-v}
\big(\det(1-K^+_v) - \det(1-K^-_v)\big)\,.
\end{eqnarray}
Here the determinant is understood in $L^2([s,\infty))$ Note that $K^\pm_v\to 0$ as $v\to 0$,  since $\Re(\eta-\zeta)>0$.
Hence for small $v$ the integral in (\ref{4.10}) is well-defined.

Next we want to reexpress the kernels $K^\pm_v$ as Airy-like
kernels. Note that, since $\Gamma^\ast_\zeta=\Gamma_\zeta$ and
$\Gamma^\ast_\eta=\Gamma_\eta$, one has
\begin{equation}\label{4.12}
K^+_v(x,y)^\ast=K^-_v(x,y)\,.
\end{equation}
We also note that, for $v=1$,
\begin{equation}\label{4.13}
K^+_1(x,y)- K^-_1(x,y)= \mathrm{i}(2\pi/\gamma_t) \mathrm{Ai}(x)\mathrm{ Ai}(y)
\end{equation}
by the standard representation of the  Airy function, denoted by Ai. Hence
\begin{equation}\label{4.14}
K^\pm_1(x,y)= B_t(x,y)\pm \mathrm{i}(\pi/\gamma_t) \mathrm{Ai}(x)\mathrm{ Ai}(y)\,,
\end{equation}
where $B_t(x,y)$ is a real kernel and defined through
\begin{equation}\label{4.15}
B_t(x,y)=\int_{\Gamma_\zeta} \mathrm{d}\zeta \int_{\Gamma_\eta} \mathrm{d}\eta (\pi/\gamma_t)
\cot (\gamma_t^{-1}\pi(\eta-\zeta))\exp\big[-\tfrac{1}{3} \zeta^3 +\tfrac{1}{3}
\eta^3 +\zeta y -\eta x\big]\,.
\end{equation}

Let $D=\partial/\partial x+\partial/\partial y$. Then
\begin{eqnarray}\label{4.16}
&&\hspace{-46pt}\int_{\Gamma_\zeta} \mathrm{d}\zeta \int_{\Gamma_\eta} \mathrm{d}\eta
(\eta-\zeta)\exp\big[-\tfrac{1}{3} \zeta^3 +\tfrac{1}{3} \eta^3 +\zeta y
-\eta x\big]\nonumber\\
&&\hspace{16pt} = -D \int_{\Gamma_\zeta} \mathrm{d}\zeta \int_{\Gamma_\eta}
\mathrm{d}\eta \exp\big[-\tfrac{1}{3} \zeta^3 +\tfrac{1}{3} \eta^3 +\zeta y -\eta
x\big] \nonumber\\
&&\hspace{16pt}=-\int \mathrm{d}\lambda \delta(\lambda) \frac{d}{d\lambda}
\mathrm{Ai}(x+\lambda)\mathrm{ Ai}(y+\lambda)\,.
\end{eqnarray}
By functional calculus this identity can be extended to a general
class of functions of $D$. To obtain the cotangent, a convenient
representation is the power series
\begin{equation}\label{4.17}
\gamma_t^{-1}\pi\cot (\gamma_t^{-1}\pi z)=\frac{1}{z} +2z \sum^\infty_{n=1}
\frac{1}{z^2-(\gamma_t n)^2}\,.
\end{equation}
The sum converges  except for  $z \in(\gamma_t\mathbb{Z})\backslash\{0\}$. Then
\begin{equation}\label{4.18}
B_t(x,y)= K_{\mathrm{Ai}}(x,y) +\int \mathrm{d}\lambda'
G(0,\lambda')\mathrm{Ai}(x+\lambda')\mathrm{ Ai}(y+\lambda')
\end{equation}
with $K_\mathrm{Ai}$ the Airy kernel
\begin{equation}\label{4.19}
K_\mathrm{Ai}(x,y)=\int^\infty_0
\mathrm{d}\lambda\mathrm{Ai}(x+\lambda)\mathrm{ Ai}(y+\lambda)
\end{equation}
and $G(\lambda,\lambda')$ the kernel of the operator
\begin{equation}\label{4.19a}
2 \frac{d}{d\lambda}\Big(\sum^\infty_{n=1}
\big(-\frac{d^2}{d\lambda^2}+(\gamma_t n)^2\big)^{-1}\Big)\,.
\end{equation}
By  direct computation one verifies that, for $a > 0$,  the operator
$2(d/d\lambda)(-(d^2/d\lambda^2) + a^2)^{-1}$ has the kernel
\begin{equation}\label{4.19b}
G_a(\lambda,\lambda') = 
- \theta
(\lambda -\lambda')\mathrm{e}^{-a|\lambda-\lambda'|}\,,
\end{equation}
where $\theta(\lambda) =-1$ for $\lambda<0$ and $ \theta(\lambda) =1$ for $\lambda \geq 0$.
Summing over $n$ as in (\ref{4.19a}) yields
\begin{equation}\label{4.20}
G(\lambda,\lambda')=- \theta
(\lambda -\lambda')(\mathrm{e}^{\gamma_t|\lambda-\lambda'|}-1)^{-1}\,.
\end{equation}
Hence one arrives at
\begin{eqnarray}\label{4.21}
&&\hspace{-26pt}B_t(x,y)
\\&&\hspace{-20pt}
= K_\mathrm{Ai}(x,y) +\int^\infty_0
\mathrm{d}\lambda(\mathrm{e}^{\gamma_t\lambda}
-1)^{-1}\big(\mathrm{Ai}(x+\lambda)\mathrm{
Ai}(y+\lambda)-\mathrm{Ai}(x-\lambda)\mathrm{
Ai}(y-\lambda)\big)\,.\nonumber
\medskip\end{eqnarray}
\textbf{Lemma 4.1.} \textit{Let $B$ be a trace class operator on the
Hilbert space $\mathcal{H}$ with scalar product $\langle\cdot,\cdot\rangle$
and $P_\psi$ the
unnormalized projection along $\psi\in\mathcal{H}$. Then for
$\alpha\in\mathbb{R}$}
\begin{eqnarray}\label{4.22}
&&\hspace{-32pt}\frac{1}{2\mathrm{i}}\big(\det(1-B-\mathrm{i}\alpha P_\psi)
-\det(1-B+\mathrm{i}\alpha P_\psi)\big)\nonumber\\
&&\hspace{6pt}
=-\alpha\big(\det(1-B+P_\psi)-\det(1-B)\big)\,.\medskip
\end{eqnarray}
\textbf{Proof.} Let $(1-B)$ be invertible. Then
\begin{equation}\label{4.23}
\det(1-B+\mathrm{i}\alpha P_\psi)=
\det(1-B)(1+\mathrm{i}\alpha\langle\psi,(1-B)^{-1}\psi\rangle)\,.
\end{equation}
Hence the difference on the left side of (\ref{4.22}) reads
\begin{equation}\label{4.24}
-\alpha\det(1-B)(1+\langle\psi,(1-B)^{-1}\psi\rangle-1)=
-\alpha(\det(1-B+P_\psi)-\det(1-B))\,.
\end{equation}
Taking limits on both sides the invertibility condition is removed.\medskip
\hfill$\Box$

To return to (\ref{4.10}) we substitute $\gamma_t u=-\log v$, 
$\mathrm{d}u=-(\gamma_t v)^{-1} \mathrm{d}v$. Then
\begin{equation}\label{4.26}
K^\pm_v(x,y)= B_t(x+u,y+u)\pm \mathrm{i}(\pi/\gamma_t) \mathrm{Ai}(x+u)\mathrm{ Ai}(y+u)\,.
\end{equation}
Using (\ref{4.22}) with $\alpha=\pi/\gamma_t$,
one arrives at the final result
\begin{eqnarray}\label{4.25}
&&\hspace{-36pt}F_t(s)= 1-\int^\infty_{-\infty}  \exp
[-\mathrm{e}^{\gamma_t(s-u)}]\nonumber\\
&&\hspace{16pt} \times
\big(\det(1-P_u(B_t-P_\mathrm{Ai})P_u)
-\det(1-P_u B_tP_u)\big)\mathrm{d}u\,.
\end{eqnarray}
Here $P_u$ projects onto $[u,\infty)$, $P_\mathrm{Ai}$ has kernel
$\mathrm{Ai}(x)\mathrm{Ai}(y)$, and the determinants are in
$L^2(\mathbb{R})$. In Appendix C we show that $B_t$ is trace class and that the integral in (\ref{4.25}) is well-defined.


\section{One-point distribution of the particle current}\label{sec5}
\setcounter{equation}{0}

Let us first state our main result concisely. We consider the WASEP with asymmetry $\beta\sqrt{\varepsilon}$ 
and have established that
\begin{equation}\label{5.1a}
\lim_{\varepsilon \to 0} \mathbb{P}\big(x_m(\varepsilon^{-2}t)
-c_1\varepsilon^{-3/2}-(c_2/\gamma_t)\varepsilon^{-1/2}
\log(2\beta\sqrt{\varepsilon})\leq c_2 s \varepsilon^{-1/2}\big) = F_t(s)\,,
\end{equation}
where
\begin{eqnarray}\label{5.1b}
&&\hspace{-20pt}m=\sigma \beta t \varepsilon^{-3/2}\,,\quad c_1=(-1+2\sqrt{\sigma})\beta t
\,,\quad c_2=\sigma^{-1/6}(1-\sqrt{\sigma})^{2/3}(\beta t)^{1/3}\,,\nonumber\\[1ex]
&&\hspace{-20pt}
\gamma_t=2\beta(\beta
t)^{1/3}(\sqrt{\sigma}(1-\sqrt{\sigma}))^{2/3}\,,\quad 0<\sigma<1\,.
\end{eqnarray}
The limit distribution function $F_t(s)$ is defined in (\ref{4.25}) and
its $t$-dependence is only through the parameter $\gamma_t$.

We reexpress our findings in terms of time-integrated currents and set \medskip\\
$\mathcal{J}^{\varepsilon}(j,t)=\sharp$ of signed jumps across the bond $(j,j+1)$ up to time
$t$,  \medskip\\
where the superscript $\varepsilon$ reminds on the
$\varepsilon$-dependence of the asymmetry. The transformation from
particle position to integrated current is discussed in \cite{TW2},
which can simply be followed (their quantity $\mathcal{I}$ equals
$-\mathcal{J}$). We also allow for a shift of the reference point by
$\varepsilon^{-1}x$, $x=\mathcal{O}(1)$. Then, for $|y|<1$, we
define the $(x,t)$-dependent family of distribution functions
\begin{eqnarray}\label{5.1c}
&&\hspace{-42pt}F^\varepsilon_{(x,t)}(s)=
\mathbb{P}\big(2\beta\sqrt{\varepsilon}\mathcal{J}^{\varepsilon}(\lfloor
y\beta t\varepsilon^{-3/2}+x\varepsilon^{-1}\rfloor,\varepsilon^{-2}
t)+\tfrac{1}{2}(1-|y|)^2
\beta^2 t \varepsilon^{-1}\nonumber\\
&&\hspace{20pt}- (1- |y|)|x|\beta\varepsilon^{-1/2}+(x^2/2 t) -\log(2
\beta\sqrt{\varepsilon}) \leq \gamma_t s\big)
\end{eqnarray}
with $\lfloor\cdot\rfloor$ denoting integer part. In the new parameters 
\begin{equation}\label{5.3}
\gamma_t=2^{-1/3} (1-y^2)^{2/3} \beta^{4/3} t^{1/3}\,.
\end{equation}
$\frac{d}{ds}F^\varepsilon_{(x,t)}(s)$ is the properly scaled
probability distribution for the statistics of the 
current at location $\lfloor y\beta t
\varepsilon^{-3/2}+x\varepsilon^{-1}\rfloor$ integrated over the time span
$[0,\varepsilon^{-2}t]$. The limit (\ref{5.1a})
can then be restated as
\begin{equation}\label{5.1d}
\lim_{\varepsilon\to 0} F^\varepsilon_{(x,t)}(s)= F_t(s)\,.
\end{equation}

Let $\xi_t$ be a random variable with distribution function $F_t$.
Then (\ref{5.1c}) translates to the one-point
statistics of the time-integrated current as
\begin{eqnarray}\label{4.28}
&&\hspace{-36pt}2\beta\sqrt{\varepsilon}\mathcal{J}^{\varepsilon}(\lfloor
y\beta
t\varepsilon^{-3/2}+x\varepsilon^{-1}\rfloor,\varepsilon^{-2} t) \nonumber\\[2ex]
&&\hspace{-20pt}\cong -\tfrac{1}{2}(1-|y|)^2 \beta^2 t
\varepsilon^{-1} + (1- |y|)|x|\beta\varepsilon^{-1/2}-(x^2/2t) + \log(2
\beta\sqrt{\varepsilon}) + \gamma_t \xi_t \,,
\end{eqnarray}
valid in the limit of small $\varepsilon$.

It is instructive to write down the probability density for $\xi_t$, which is given by
\begin{equation}\label{4.27}
\rho_t (s)= \frac{d}{ds}F_t(s)= \int^\infty_{-\infty}  \gamma_t
\mathrm{e}^{\gamma_t(s-u)}\exp \big[-\mathrm{e}^{\gamma_t(s-u)}\big]g_t(u)\mathrm{d}u\,,
\end{equation}
where
\begin{equation}\label{4.27a}
g_t(u) = \det (1-P_u
(B_t-P_\mathrm{Ai})P_u) -\det(1-P_u
B_tP_u)\,.
\end{equation}
$1-  \exp
[-\mathrm{e}^{\gamma_ts}]$ is the Gumbel distribution function and the first factor
in (\ref{4.27}) is the Gumbel probability density. Since
$\lim_{s\to-\infty}F_t(s)=0$, it follows that the second factor of the convolution is normalized as
\begin{equation}\label{4.26a}
\int^\infty_{-\infty}
g_t(u)\mathrm{d}u=1\,.
\end{equation}
However, numerical computations clearly indicate that the definite sign of $\rho_t$ is regained
only after smearing with the Gumbel density, see \cite{SS}.

As discussed in the Introduction, in the long time limit the integrated current
is expected to be Tracy-Widom distributed, which corresponds to taking
$t \to \infty$ in (\ref{4.28}). Since $\gamma_t\sim
t^{1/3}$,  the Gumbel distribution in (\ref{4.27}) tends to $\delta(s-u)$ and $B_t$ 
of (\ref{4.21}) tends to $K_\mathrm{Ai}$ as $t\to\infty$.
Hence $\xi_\infty$ has the probability density
\begin{equation}\label{4.28a}
\rho_\infty(s)=g_\infty(s)=F_\mathrm{TW}(s) u(s)
\end{equation}
with
\begin{equation}\label{4.29}
F_\mathrm{TW}(s) = \det (1-P_s K_\mathrm{Ai}P_s)\,,\quad
u(s)=\langle P_s \mathrm{Ai},(1-P_s K_\mathrm{Ai}P_s)^{-1} P_s
\mathrm{Ai}\rangle\,.
\end{equation}
It follows from the identities in \cite{TW1} that
\begin{equation}\label{4.30}
(\log F_\mathrm{TW})'=u\,.
\end{equation}
Hence $\rho_\infty=F'_\mathrm{TW}$ and
\begin{equation}\label{4.31}
\lim_{t\to\infty} F_t(s)=F_\mathrm{TW} (s)\,,
\end{equation}
confirming the conventional expectation.

\section{Conclusions}\label{sec6}
\setcounter{equation}{0}

We have studied the crossover asymptotics of the WASEP in an $\varepsilon$-dependent
parameter window. Somewhat more physically,
our result can be rephrased through appropriately adjusted
time scales. One considers the asymmetry $q-p$ which is assumed to
be small but fixed. $(q-p)^{-1}$ defines the time unit. For short
times the statistics  of the integrated current is approximately
Gaussian. In an intermediate time window one should observe the
statistics corresponding to $F_t$, which for long times approaches the
Tracy-Widom distribution $F_\mathrm{TW}$. As $q-p$ is increased the crossover
window should shrink and might become not discernible at all.

In this paper the focus is on the derivation of $F_t(s)$ from the
WASEP scaling limit. Properties of this family of distribution
functions  and their relation to the KPZ equation will be discussed in the companion papers
\cite{SS,SS1}.\medskip\\
\textit{Remark}: After submitting  this article, G. Amir, I. Corwin, and J. Quastel posted their paper \cite{ACQ}
in the arXiv, in which independently  they
establish the limit (\ref{5.1a}) and the formula (\ref{4.27}) for the probability 
density $\rho_t$. As here, their starting point is the PASEP Tracy-Widom contour integration formula.
\bigskip\\
\textbf{Acknowledgements.} We are grateful to Michael Pr\"{a}hofer
for many illuminating discussions and to Sylvain Prolhac for helping with the combinatorics in Appendix C.
H.~S. thanks Jeremy
Quastel for emphasizing the importance of the crossover WASEP.
This work is supported by a DFG grant. In addition T.S. acknowledges the support from
KAKENHI (9740044) and H.S. from Math-for-Industry
of Kyushu University.


\begin{appendix}
\section{Appendix: The set $\mathcal{D}_\kappa$}\label{app.B}
\setcounter{equation}{0}

For $\kappa >0$ we define
$\mathcal{D}_\kappa=\mathcal{D}^{(1)}\cap\mathcal{D}^{(2)}_\kappa$,
with
\begin{equation}\label{B.1}
\mathcal{D}^{(1)}=\big\{\mu\big|\sup_{0\leq y\leq 1}\big|\frac{\mu y}{1-\mu
y}\big|\leq 1\big\}\,,\;
\mathcal{D}^{(2)}_\kappa=\big\{\mu\big|\sup_{0\leq y\leq 1}\big|\frac{y}{\mu-
y}\big|\leq1+ \kappa\big\}\,,
\end{equation}
and $\Gamma_\kappa = \{ \mu | \mathrm{dist} (\mu,\mathbb{R}_{+}) = 1, \Re
\mu <\kappa \}$. We claim that one can
choose $\varepsilon_0>0$ such that for all
$0<\varepsilon<\varepsilon_0$ it holds
\begin{equation}\label{B.1a}
2\beta\sqrt{\varepsilon}\,\Gamma_\kappa\subset\mathcal{D}_\kappa\,. 
\end{equation}

For $\mu \in 2\beta\sqrt{\varepsilon}\,\Gamma_\kappa$ one has $|\Re
\mu| <2\beta(1 +\kappa)\sqrt{\varepsilon}$, $|\Im
\mu| <2\beta\sqrt{\varepsilon}$. Clearly, for sufficiently small $\varepsilon$
our assertion holds for the set $\mathcal{D}^{(1)}$.

To discuss $\mathcal{D}^{(2)}_\kappa$ we define
\begin{equation}\label{B.3}
\ell(y)=y\big((\tilde{a}-y)^2 +\tilde{b}^2\big)^{-1/2}\,,\quad 0\leq y\leq 1\,.
\end{equation}
(i) For $\tilde{a}<0$, the maximum of $\ell$ is at $y_0=1$ and it holds that $\ell\leq1$. (ii) For
$\tilde{a}>0$ and $\tilde{a}^2+\tilde{b}^2<\tilde{a}$ the maximum of $\ell$ 
is at $y_0=(\tilde{a}^2+\tilde{b}^2)/\tilde{a}$ 
and $\ell\leq |\tilde{b}|^{-1}(\tilde{a}^2+\tilde{b}^2)^{1/2}$. For 
$\mu \in 2\beta\sqrt{\varepsilon}\,\Gamma_\kappa$ we set $\mu = 2\beta\sqrt{\varepsilon}(a+ib)$. If 
$a<0$, we are in case (i) and our assertion holds for this part of the contour. If $a\geq 0$, then 
$\tilde{a} = 2\beta\sqrt{\varepsilon}a$, $ a <\kappa$, $\tilde{b} = 2\beta\sqrt{\varepsilon}b$, $|b| =1$. For $\varepsilon$
sufficiently small it holds $2\beta\sqrt{\varepsilon}(a^2 + b^2) < a$ and we are in case (ii).
Hence $\ell$ is bounded by $b^{-1}(a^2+b^2)^{1/2} \leq (\kappa^2 +1)^{1/2} \leq 1+ \kappa$,
which implies that also this part of the contour is contained in $\mathcal{D}^{(2)}_\kappa$.


\section{Appendix: Bounds on the kernels $I(\mu),K^\pm_v$}\label{app.A}
\setcounter{equation}{0}

The $\zeta$-integration in $I(\mu)$ from (\ref{3.24}) has a simple
pole whenever
\begin{equation}\label{A.1}
\eta'-\zeta=\gamma_t m\,,\quad m\in\mathbb{Z}\,.
\end{equation}
Thus the $\zeta$-integration is singular for a discrete set of
points $\eta'\in\Gamma_\eta$. The integration along $\Gamma_\zeta$ looks
locally like
\begin{equation}\label{A.2}
\int^a_{-a} \mathrm{d}u(u-z)^{-1}
\end{equation}
with sufficiently small $a>0$. For $z\neq 0$ and $|z|<a/2$, the integral
(\ref{A.2}) is uniformly bounded and extends by continuity to $z=0$.
Thus $|I(\mu;\eta,\eta')|$
is bounded by $c_0e^{-\delta|\eta|^3}e^{-\delta|\eta'|^3}$ for
sufficiently small $\delta>0$. Hence $I(\mu)$ is trace class.

A similar discussion applies to the kernels $K^\pm_v (x,y)$ from
(\ref{4.11}). The singularities of $\csc(\pi \gamma_t^{-1}(\eta-\zeta))$ are
simple isolated poles only, hence integrable, the integrations in (\ref{4.11})
are well-defined, and the kernel is bounded by $c_0e^{-\delta|\eta|^3}e^{-\delta|\eta'|^3}$ for
sufficiently small $\delta>0$.

\section{Appendix: $\mu$-integration first}\label{C}
\setcounter{equation}{0}

We describe an alternative route to arrive at (\ref{4.10}). The starting
point is the expansion of the Fredholm determinant (\ref{2.1}), which is a
$2n$-fold integral over $\zeta_1,\ldots,\zeta_n$ and
$\eta_1,\ldots,\eta_n$. Each integrand has factors corresponding to
$Q_1,Q_2,Q_3$ as in (\ref{3.9}). For each of them we use the saddle point
approximation as discussed in Section \ref{sec3}. There then remains an
$n$-fold product of functions $f$ with an argument scaled as in
(\ref{3.9}), (\ref{3.9b}). Finally the $\mu$-integration has to be carried out,
see (\ref{2.1}).

The new idea here is to first integrate over $\mu$ and then take the
limit $\varepsilon\to 0$ with the arguments $\sqrt{\varepsilon}$
close to the saddle. Thereby one circumvents the discussion at the
end of Section \ref{sec3}. With a more careful variant, possibly, one could
control the error bounds.

To lighten the notation, we use instead of $\tau$ the conventional symbol $q = 
1- \sqrt{\varepsilon}$. For $1<|z_j|<q^{-1}$, $j=1,\ldots,n$, we define
\begin{eqnarray}\label{C.1}
&&\hspace{-58pt}H_n(z_1,\ldots,z_n)=\int_{\mathcal{C}_0}\mathrm{d}\mu
\frac{1}{\mu} g_q(\mu)
\prod^n_{j=1}  \{\mu f(\mu, z_j)\}\nonumber\\
&&\hspace{20pt}=(-1)^n\sum_{(\ell_1,\ldots,\ell_n)\in\mathbb{Z}^n}
\int_{\mathcal{C}_0}\mathrm{d}\mu \mu^{n-1} g_q(\mu) \prod^n_{j=1}
\{(\mu - q^{-\ell_j})^{-1}(z_j)^{\ell_j}\}\,,
\end{eqnarray}
with
\begin{equation}\label{C.2}
g_q(\mu)=\prod^\infty_{k=0}(1-\mu q^k)\,.
\end{equation}
We split the sum over $\ell_1,\ldots,\ell_n$ into $n$-tuples
with no double points and the rest. Thereby 
\begin{equation}\label{C.3}
H_n=H_n^{(\ast)}+H_n^{\textrm{rem}}\,.
\end{equation}
Only $H_n^{(\ast)}$ is discussed.  We checked that the remainder term $H_n^{\textrm{rest}}$ 
tends to $0$ for $\varepsilon \to 0$ for $n=2,3,4,5$. Unfortunately the combinatorial 
structure becomes involved and we did not try to work out the extension to general $n$.

$H_n^{(\ast)}$ has only the simple poles $q^{-\ell}$,
$\ell\in\mathbb{Z}$. $\mathcal{C}_0$ encloses those poles with
$\ell<0$. Therefore the $\mu$-integration can be carried out with
the result
\begin{eqnarray}\label{C.4}
&&\hspace{-62pt}H^{(\ast)}_n(z_1,\ldots,z_n)= (-1)^n \sum_{\hspace{10pt}(\ell_1,\ldots,\ell_n)\in\mathbb{Z}^n}
\hspace{-20pt}{^{(\ast)}}\hspace{10pt}
\sum^n_{i=1}\chi(\{\ell_i<0\}) q^{-(n-1)\ell_i} g_q(q^{-\ell_i})\nonumber\\
&&\hspace{34pt}  \times\prod^n_{j=1}{^{(i)}}
(q^{-\ell_i}-q^{-\ell_j})\prod^n_{j=1} (z_j)^{\ell_j}\,.
\end{eqnarray}
Here $\sum^{(\ast)}$ means no double points and $\prod^{(i)}$ means
that the $i$-th factor is omitted from the product. We now substitute $ \ell_j$ by
$\ell_j+\ell_i$ for $j\neq i$. Then
\begin{eqnarray}\label{C.5}
&&\hspace{-62pt} H^{(\ast)}_n(z_1,\ldots,z_n)=
(-1)^n \Big(\sum^\infty_{\ell>0}
g_q(q^\ell)\prod^n_{j=1} (z_j)^\ell\Big)\nonumber\\
&&\hspace{38pt}\times\sum^n_{i=1}\sum_{\hspace*{6pt}(\ell_1,\ldots,\widehat{\ell}_i,\ldots,\ell_n)\in(\mathbb{Z}\setminus
0)^{n-1}} \hspace*{-40pt}{^{(\ast)}} \hspace*{30pt}\prod^n_{j=1}\hspace{-2pt}{^{(i)}}\hspace{2pt}
\{(1-q^{-\ell_j})^{-1} (z_j)^{\ell_j}\}\nonumber\\
&&\hspace{22pt}=H^{(\ast),\Gamma}_n(z_1,\ldots,z_n)
H^{(\ast),\Sigma}_n(z_1,\ldots,z_n)\,.
\end{eqnarray}

We first study the issue of analytic extension. Since $g_q(q^{\ell})\cong
\exp[-c q^\ell]$, $H^{(\ast),\Gamma}_n$ is analytic on all
of $\mathbb{C}^n$. For $H^{(\ast),\Sigma}_n$ it suffices to consider
the case $i=n$. From the  restriction of no double points, we conclude
that we have a sum over products, where each factor is of the form
\begin{eqnarray}\label{C.6}
&&\hspace{-12pt}\sum_{m\neq 0}(1-q^{-m})^{-k} z^m
=\sum^\infty_{m=1}\big\{(-1)^k(1-q^m)^{-k} (q^k z)^m +  (1-q^m)^{-k}
z^{-m}\big\}\\
&&\hspace{4pt}=\sum^{\infty}_{\ell_1,\ldots,\ell_n=0}\,\,\sum^{\infty}_{m=1}\big\{
(-1)^k q^{m(\ell_1+\ldots+\ell_k)}(q^k z)^m + 
q^{m(\ell_1+\ldots+\ell_k)} z^{-m}\big\}\nonumber\\
&&\hspace{4pt}=\sum^{\infty}_{\ell_1,\ldots,\ell_k=0} \big\{ (-1)^k (1-z
q^k q^{\ell_1+\ldots+\ell_k})^{-1} z q^k q^{\ell_1+\ldots+\ell_k} +
(z-q^{\ell_1+\ldots+\ell_k})^{-1} q^{\ell_1+\ldots+\ell_k}
\big\}\nonumber
\end{eqnarray}
for $k=1,2,\ldots$. The last expression is the analytic extension to
$\mathbb{C}\setminus\{0\}$ except for the poles at $z=q^\ell$,
$\ell\in\mathbb{Z}$.

We summarize the result in\medskip\\
\textbf{Proposition C.1}. \textit{Let $\mathcal{D}_a=\{z\in\mathbb{C}\mid
1<|z|<q^{-1}\}$ and let $\mathcal{D}_q=\{z\in\mathbb{Z}\mid z\neq 0,
z\neq q^\ell, \ell\in\mathbb{Z}\}$. Then $H^{(\ast)}_n$, as defined
on $\mathcal{D}^{\otimes n}_a$, extends to an analytic function on
$\mathcal{D}^{\otimes n}_q$.}\medskip

Close to the saddle $z_j=1+\sqrt{\varepsilon}w_j$. For
$1+\sqrt{\varepsilon}w_j\in\mathcal{D}_q$ we define
\begin{eqnarray}\label{C.7a,C.7b,C.7c}
&&\hspace{-16pt}H^{(\ast)}_{n,\varepsilon}=H^{(\ast),\Gamma}_{n,\varepsilon}
H^{(\ast),\Sigma}_{n,\varepsilon}\,,\\[1ex]
&&\hspace{-16pt}H^{(\ast),\Gamma}_{n,\varepsilon}(w_1,\ldots,w_n)=(-1)^n(\sqrt{\varepsilon})^{1-w}
H^{(\ast),\Gamma}_n(1+\sqrt{\varepsilon}
w_1,\ldots,1+\sqrt{\varepsilon} w_n)\,,\\[1ex]
&&\hspace{-16pt}H^{(\ast),\Sigma}_{n,\varepsilon}(w_1,\ldots,w_n)=(\sqrt{\varepsilon})^{n-1}
H^{(\ast),\Sigma}_n(1+\sqrt{\varepsilon}
w_1,\ldots,1+\sqrt{\varepsilon} w_n)\,,
\end{eqnarray}
with the shorthand $w=\sum^n_{j=1} w_j$. We also introduce the limit
functions defined on $\mathcal{D}^{\otimes n}_0$, with
$\mathcal{D}_0=\mathbb{C}\setminus\mathbb{Z}$,
\begin{equation}\label{C.8a}
H^\Gamma_n (w_1,\ldots,w_n)= (-1)^n\Gamma (w)\,,
\end{equation}
\begin{equation}\label{C.8b}
H^\Sigma_n (w_1,\ldots,w_n)= \pi^{-1}\sin(\pi
w)\Big(\prod^n_{j=1} \pi^{-1}\sin(\pi w_j)\Big)^{-1}\,.
\medskip\end{equation}
\textbf{Theorem C.2.} 
\textit{Pointwise on $\mathcal{D}^{\otimes n}_0$ it
holds}
\begin{equation}\label{C.9}
\lim_{\varepsilon\to 0} H^{(\ast),\Gamma}_{n,\varepsilon} =
H^\Gamma_n\,, \quad \lim_{\varepsilon\to 0}
H^{(\ast),\Sigma}_{n,\varepsilon}=H^\Sigma_n\,.
\medskip\end{equation}
The proof is divided into several parts. We start with $g_q$.\medskip\\
\textbf{Lemma C.3.} \textit{
It holds}
\begin{equation}\label{C.10}
\lim_{\varepsilon\to 0} (\sqrt{\varepsilon})^{-w+1} \sum^\infty_{\ell=1}
g_q(q^\ell)
\big(\prod^n_{j=1} (1+\sqrt{\varepsilon} w_j)\big)^{-\ell}=\Gamma(w)\,.\medskip
\end{equation}
\textbf{Proof.} 
The product in (\ref{C.10}) equals
$1+\sqrt{\varepsilon}w+\mathcal{O}(\varepsilon^2)$. Then, by \cite{A},
Eq. 10.3.1,
\begin{eqnarray}\label{C.11}
&&\hspace{-50pt}(\sqrt{\varepsilon})^{-w+1}
\sum^\infty_{\ell=1}
g_q(q^\ell)
(1+\sqrt{\varepsilon}w)^{-\ell}=(\sqrt{\varepsilon})^{-w+1}\sum^{\infty}_{\ell=0}
(q^{\ell+1};q)_{\infty}(1+\sqrt{\varepsilon}w)^{-\ell-1}\nonumber\\
&&\hspace{46pt}=(\sqrt{\varepsilon})^{-w+1}
(1+\sqrt{\varepsilon}w)^{-1}\frac{(q;q)_\infty}{(y;q)_\infty}\,,\quad
y=(1+\sqrt{\varepsilon}w)^{-1}\,.
\end{eqnarray}
The $q$-Gamma function is defined by
\begin{equation}\label{C.12}
\Gamma_q(x)=
(1-q)^{1-x}\frac{(q;q)_\infty}{(q^x;q)_\infty}\,,\quad |q|<1\,.
\end{equation}
Setting $x=w$, one thus arrives at
\begin{equation}\label{C.13}
(\sqrt{\varepsilon})^{-w+1}
\sum^\infty_{\ell=1}
g_q(q^\ell)
(1+\sqrt{\varepsilon}w)^{-\ell}=(1+\sqrt{\varepsilon}w)^{-1}\Gamma_q(w)\,,
\end{equation}
which implies the limit (\ref{C.10}).\hfill$\Box$\medskip

Lemma 3 establishes the left part of (\ref{C.9}).\medskip\\
\textbf{Lemma C.4.} \textit{The following limits hold.\\
(i) For $k = 1$
\begin{eqnarray}\label{C.14}
&&\hspace{-18pt}
\lim_{\varepsilon\to 0}
\sqrt{\varepsilon}\sum^{\infty}_{\ell=0}\big\{-\big(1-(1+\sqrt{\varepsilon}w_1)q^{\ell+1}\big)^{-1}
(1+\sqrt{\varepsilon}w_1) q^{\ell+1}+\big((1+\sqrt{\varepsilon}w_1)-
q^{\ell}\big)^{-1} q^\ell\big\}\nonumber\\
&&\hspace{32pt}= \pi \cot \pi w_1\,.
\end{eqnarray}
\textit{(ii)} For} $k\geq 2$
\begin{eqnarray}\label{C.15}
&&\hspace{-36pt}\lim_{\varepsilon\to 0}
(\sqrt{\varepsilon})^k\sum^{\infty}_{\ell_1,\ldots,\ell_k=0}
\big\{(-1)^k\big(1-(1+\sqrt{\varepsilon}w_1)q^{k+\ell_1+\ldots+\ell_k}\big)^{-1}
(1+\sqrt{\varepsilon}w_1) q^{k+\ell_1+\ldots+\ell_k}\nonumber\\
&&\hspace{46pt}+\big((1+\sqrt{\varepsilon}w_1)
-q^{\ell_1+\ldots+\ell_k}\big)^{-1}q^{\ell_1+\ldots+\ell_k}\big\}\nonumber\\
&&\hspace{12pt}=\big(1+(-1)^k\big) \frac{1}{(k-1)!}\int^\infty_0
\mathrm{d}u u^{k-1} (\mathrm{e}^u-1)^{-1}=\frac{1}{(k-1)!}B(k)\,.\medskip
\end{eqnarray}

For even $k$ the coefficients $B(k)$ are related to the Bernoulli
numbers $B_k$ by $B(k)=k^{-1} (2\pi)^k (-1)^{1+(k/2)}
B_k$.\medskip\\
\textbf{Proof.} \textit{ad (i):} Separating the $\ell=0$ term of
the second summand and expanding inside the curly bracket one
obtains
\begin{eqnarray}\label{C.16}
&&\hspace{-16pt}\lim_{\varepsilon\to 0} \sqrt{\varepsilon}
\big[-\big(1-(1+\sqrt{\varepsilon}w_1)\big)^{-1} -
\sum^\infty_{\ell=1}\big\{(1+\sqrt{\varepsilon}w_1)(1-\sqrt{\varepsilon}\ell)
\big(1-(1+\sqrt{\varepsilon}w_1) \nonumber\\
&&\hspace{44pt}\times(1-\sqrt{\varepsilon}\ell)\big)^{-1} - (1-\sqrt{\varepsilon}w_1)(1-\sqrt{\varepsilon}\ell)
\big(1-(1-\sqrt{\varepsilon}w_1) (1-\sqrt{\varepsilon}\ell)\big)^{-1}\big\}\big]\nonumber\\
&&\hspace{30pt}= (w_1)^{-1}-\sum^\infty_{\ell=1}2
w_1(w^2_1-\ell^2)^{-1}\nonumber\\
&&\hspace{30pt}=\pi \cot \pi w_1\,.
\end{eqnarray}
\textit{ad (ii):} For $k\geq 2$ the sum approximates the Riemann
integral
\begin{equation}\label{C.16a}
\int^\infty_0 \mathrm{d} u_1 \ldots \int^\infty_0 \mathrm{d} u_k
\prod^k_{j=1} \mathrm{e}^{-u_j} \big(1-\prod^k_{j=1}
\mathrm{e}^{-u_j}\big)^{-1} \big(1+(-1)^k\big)\,.
\end{equation}
Note that the limit does not depend on  $w_1,\ldots,w_k$, in
contrast to $k=1$.\hfill$\Box$\medskip

With Lemma 3 and 4 we have identified the limit of
$H^{(\ast),\Sigma}_{n,\varepsilon}$, in principle. It remains for
each $n$ to rearrange the sum such that the expression (\ref{C.8b})
results.

To handle the constraint of the sum in (\ref{C.5}), let us denote by
$I_{n-1}$ the set $\{1,\ldots,n-1\}$ of labeled vertices. An
unoriented edge with endpoints $i,j$, $i\neq j$, is denoted by $b$
and $\delta_b$ stands for $\delta_{\ell_i\ell_j}$. Let
$\mathcal{E}_{n-1}$ be the set of all edges of $I_{n-1}$. Then the
constraint of the sum in (\ref{C.5}) can be written as
\begin{equation}\label{C.17}
\prod_{b\in\mathcal{E}_{n-1}}(1-\delta_b)\,.
\end{equation}
By expanding the product one has to sum over the set
$\mathcal{G}_{n-1}$ of all undirected graphs over $I_{n-1}$. For
$g\in\mathcal{G}_{n-1}$ the weight, $w(g)$, in this sum results from
$(-1)^{\sharp(\textrm{edges})}$ and from the limits (\ref{C.14}) and 
(\ref{C.15}). A given graph $g$ decomposes $I_{n-1}$ into disjoint
clusters. In view of (\ref{C.8b}) and (\ref{C.14}) we keep the
number of clusters of size 1 fixed (they yield a product of
cotangents) and sum over all other clusters, by (\ref{C.15})
necessarily of even size. Thereby we arrive at the following
counting problem.

Given is the set $I_m$ of $m$ vertices, $m$ even, and set
$\mathcal{G}_m$ of undirected graphs over $I_m$. For a given graph
$g\in\mathcal{G}_m$, $I_m$ decomposes into $r$ clusters
$C_1,\ldots,C_r$ of size $|C_j|=m_j$, $\sum^r_{j=1} m_j=m$. The
weight $w(g)$ is defined by
\begin{equation}\label{C.17a}
w(g)=(-1)^{\sharp(\textrm{edges})} \prod^r_{j=1} B(m_j)
((m_j-1)!)^{-1}\,.
\end{equation}
Note that $B(m)=0$ for odd $m$. Since we do not want to allow for
$m_j=1$, only for the next lemma we set
\begin{equation}\label{C.17b}
B(1)=0\,.\medskip
\end{equation}
\textbf{Lemma C.5.} \textit{For even $m$ it holds}
\begin{equation}\label{C.18}
\sum_{g\in\mathcal{G}_m} w(g)=\frac{\pi^m}{m+1}(-1)^{m/2}\,.\medskip
\end{equation}
\textbf{Proof.} The weight $w(g)$ induces a weight of the clusters as
\begin{equation}\label{C.19}
w(\{C_1,\ldots,C_r\})=\prod^r_{j=1} w(C_j)\,.
\end{equation}
To compute $w(C_j)$ we note that
\begin{equation}\label{C.20}
\sum_{g\in\mathcal{G}_k,\,\textrm{single cluster}} w(g)=(k-1)!
(-1)^{k-1} ((k-1)!)^{-1} B(k)\,.
\end{equation}
The prefactor can be verified firstly because it holds for $k=2$.
Now assume it is valid for general $k>2$. Then, adding an extra
vertex, $k+1$, it can be connected in $k$ distinct ways to the
cluster of size $k$ and the number of edges is thereby always
increased by 1. Hence
\begin{equation}\label{C.21}
    w(C_j)=(-1)^{m_j-1} B(m_j)\,.
\end{equation}

We introduce a generating function, $f(\lambda)$, for the left hand
side of (\ref{C.18}) by
\begin{eqnarray}\label{C.22}
&&\hspace{6pt}f(\lambda)=\sum^\infty_{\textrm{even } m=2}
\frac{\lambda^m}{m!}\sum_{g\in\mathcal{G}_m} w(g)\nonumber\\
&&\hspace{32pt}=\sum^\infty_{\textrm{even } m=2}\sum^\infty_{r=1}
\frac{1}{m!}\lambda^m\frac{1}{r!}\sum_{m_1,\ldots,m_r\geq 2}\delta
(\sum^r_{j=1} m_j-m)\nonumber\\
&&\hspace{44pt}\times\left(
                   \begin{array}{c}
                     m \\
                     m_1\ldots m_r \\
                   \end{array}
                     \right)(-1)^r \prod^r_{j=1} B(m_j)
\nonumber\\
&&\hspace{32pt}=\sum^\infty_{r=1} \frac{1}{r!}(-1)^r
\big(\sum^\infty_{k=2}\frac{1}{k!} \lambda^k B(k)\big)^r\,.
\end{eqnarray}
Using (\ref{C.15}) the sum over $k$ reads
\begin{eqnarray}\label{C.23}
&&\hspace{-28pt}\sum^\infty_{n=1} \frac{1}{(2n)!}\lambda^{2n} 2
\int^\infty_0 \mathrm{d}u u^{2n-1} (\mathrm{e}^u -1)^{-1}\nonumber\\
&&\hspace{20pt}=\int^\infty_0 \mathrm{d}u (u(\mathrm{e}^u -1))^{-1}
2(\cosh \lambda u-1)=\log (\pi\lambda/\sin\pi\lambda)\,.
\end{eqnarray}
Hence
\begin{equation}\label{C.24}
f(\lambda)= (\pi\lambda)^{-1}\sin(\pi\lambda)-1\,.
\end{equation}
Taylor expanding $f$ confirms the claim.\hfill$\Box$\medskip\\
\textbf{Proof of Theorem C.2.} We return to $H^\Sigma_n$ with the
constraint in the summation written as in (\ref{C.17}). For
$g\in\mathcal{G}_{n-1}$ we decompose $I_{n-1}=I^1_{n-1} \cup
I^2_{n-1}$. $I^1_{n-1}$ consists of one point clusters $\{i\}$,
$i\in I^1_{n-1}$, and $I^2_{n-1}$ consists of clusters of size $\geq
2$, compare with the notation above (\ref{C.17a}). By (\ref{C.14})
the one point cluster $\{i\}$ carries the weight
\begin{equation}\label{C.25}
h_i=\pi \cot (\pi w_i)\,.
\end{equation}
Either set could be empty and $|I^2_{n-1}|$ is even. With this
convention the weight of $g$ is given by
\begin{equation}\label{C.26}
w(g)=(-1)^{\sharp\textrm{(edges)}}\Big(\prod_{i\in I^1_{n-1}} h_i\Big)
\prod^r_{j=1} \{B(m_j) ((m_j-1)!)^{-1}\}\,,
\end{equation}
where the second product refers to the clusters of $I^2_{n-1}$, see
(\ref{C.17a}). If $I^1_{n-1}=\emptyset$, then the first product equals 1
and correspondingly for $I^2_{n-1}$. Then
\begin{equation}\label{C.27}
H^\Sigma_n
(w_1,\ldots,w_n)=\sum^n_{i=1}f_{n-1}(w_1,\ldots,\widehat{w}_i,\ldots,w_n)
\end{equation}
with
\begin{equation}\label{C.28}
f_{n-1}(w_1,\ldots,w_{n-1})=\sum_{g\in\mathcal{G}_{n-1}} w(g)\,.
\end{equation}

It is convenient to introduce the symmetrizer $\textsf{S}_n$. If $g$
is some function on $\mathbb{C}^m$, $1\leq m\leq n$, then
$\textsf{S}_n g$ is defined by
\begin{equation}\label{C.29}
(\textsf{S}_n g)(w_1,\ldots,w_n)=\frac{1}{n!}\sum_\pi
g(w_{\pi(1)},\ldots,w_{\pi(m)})\,,
\end{equation}
where the sum is over all permutations $\pi$ of $(1,\ldots,n)$. Let
us also set $|I^2_{n-1}|=j$, $j=0,2,\ldots,[n]$, where $[n]=n-2$ for
even $n$ and $[n]=n-1$ for odd $n$. In the sum (\ref{C.28}) we fix
the set $I^1_{n-1}$ and sum over all graphs for the set $I^2_{n-1}$.
By Lemma C.5 this yields
\begin{equation}\label{C. 29a}
\Big(\prod_{i\in I^1_{n-1}} h_i\Big)\frac{1}{j+1}\pi^j(-1)^{j/2}\,.
\end{equation}
Performing the sum over all subsets $I^1_{n-1}$, one obtains
\begin{equation}\label{C.30}
f_{n-1}(w_1,\ldots,w_{k-1})=\sum^{[n]}_{j=0,j\,\textrm{even}}(-1)^{j/2}
\pi^j (j+1)^{-1} \left(
                   \begin{array}{c}
                     n-1 \\
                     n-1-j \\
                   \end{array}
                 \right)\textsf{S}_{n-1}(h_1\cdots h_{n-1-j})\,.
\end{equation}
Carrying out the sum (\ref{C.27}), it then follows that
\begin{eqnarray}\label{C.31}
&&\hspace{-28pt}H^\Sigma_n(w_1,\ldots,w_n)=\sum^n_{j=0,\,\textrm{even}}
(-1)^{j/2} \pi^j \left(
                   \begin{array}{c}
                     n \\
                     n-j \\
                   \end{array}
                 \right)\textsf{S}_{n}(h_1\cdots h_{n-1-j})\nonumber\\
&&\hspace{20pt}=\pi^{-1}\sin\big(\pi\sum^n_{j=1}w_j\big)\Big(\prod^n_{j=1}
\pi^{-1}\sin(\pi w_j)\Big)^{-1} \,.
\end{eqnarray}
The last term is a mere rewriting of the identity
\begin{equation}\label{C.32}
\sin(\sum^n_{j=1}
\theta_j)=\sum_{\substack{\textrm{odd }k\geq 1\\k\leq n}}
(-1)^{(k-1)/2} \sum_{\substack{A\subset\{1,\ldots,n\}\\|A|=k}}
\prod_{i\in A} \sin \theta_i \prod_{i\in A^c} \cos
\theta_i\,.
\end{equation}
(\ref{C.31}) establishes the second assertion of Theorem C.2.\medskip\hfill$\Box$

If the remainder term in (\ref{C.3}) is ignored, then by Theorem C.2 the scaled $H_n$ converges to
\begin{equation}
H^\Gamma_n H^{(\ast),\Sigma}_{n}=\Gamma(w) \pi^{-1}\sin\big(\pi\sum^n_{j=1}w_j\big)\Big(\prod^n_{j=1}
\pi^{-1}\sin(\pi w_j)\Big)^{-1}\,.
\end{equation}
This agrees with (\ref{4.5}) upon performing the $\mu$-integration by using the first identity of (\ref{4.6})
and collecting the factors from the saddle point.


\section{Appendix: Trace class property}\label{app.D}
\setcounter{equation}{0}

\textbf{Proposition D.1}: \textit{The operator $P_s B_t P_s$ is trace class.
The functions appearing in (\ref{4.25}) and (\ref{4.27})
are absolutely integrable in $u$.\medskip\\}
\textbf{Remark}. We have no direct proof that $\rho_t(s)\geq 0$ and
$\int^\infty_{-\infty} \rho_t (s)\mathrm{d}s=1$.\medskip\\
\begin{proof} All operators will be defined on
$L^2\big([s,\infty)\big)$.  

We have
\begin{equation}\label{D.1}
    B_t=K_\mathrm{Ai} +C_t\,,
\end{equation}
where for simplicity we set $\gamma_t=1$. In general, if $B=A_1A_2$, then
$(\mathrm{tr}|B|)^2\leq (\mathrm{tr}A^\ast_1
A_1)(\mathrm{tr}A^\ast_2 A_2)$, see \cite{RS1}, Section VI.6. For the Airy kernel we write
\begin{equation}\label{D.2}
K_\mathrm{Ai}(x,y)=\int^\infty_s \mathrm{d}\lambda
\mathrm{Ai}(x+\lambda-s)\mathrm{Ai}(y+\lambda-s)\,.
\end{equation}
Hence $K_\mathrm{Ai}=A^2_1$ with $A_1(x,y)=\mathrm{Ai}(x+y-s)$ and
\begin{equation}\label{D.3}
\mathrm{tr}|K_\mathrm{Ai}|\leq \int^\infty_s
\mathrm{d}x\int^\infty_s \mathrm{d}y |\mathrm{Ai}(x+y)|^2\,.
\end{equation}
For the operator $C_t$ we write
\begin{eqnarray}\label{D.4}
&&\hspace{-32pt}C_t(x,y)= \int^\infty_0 \mathrm{d}\lambda
(\mathrm{e}^\lambda-1)^{-1}\big(\mathrm{Ai}(x+\lambda)\mathrm{Ai}(y+\lambda)-
\mathrm{Ai}(x-\lambda)\mathrm{Ai}(y-\lambda)\big)\nonumber\\
&&\hspace{10pt}=\int^\infty_s \mathrm{d}\lambda
(\mathrm{e}^{\lambda-s}-1)^{-1}\big(\mathrm{Ai}(x+\lambda-s)-\mathrm{Ai}(x-\lambda+s)\big)
\mathrm{Ai}(y+\lambda-s)\nonumber\\
&&\hspace{22pt}+\int^\infty_s \mathrm{d}\lambda
\mathrm{e}^{-(\lambda-s)/2}
\mathrm{Ai}(x-\lambda+s)\big(\mathrm{Ai}(y+\lambda-s)-
\mathrm{Ai}(y-\lambda+s)\big)\nonumber\\
&&\hspace{62pt}\times(\mathrm{e}^{\lambda-s}-1)^{-1}\mathrm{e}^{(\lambda-s)/2}\,.
\end{eqnarray}
Hence $C_t=A_2 A_1+ A_3 A_4$.

By definition
\begin{equation}\label{D.5a}
\mathrm{tr}|A_2|^2= \int^\infty_s \mathrm{d}x
\int^\infty_0 \mathrm{d}y
(\mathrm{e}^y-1)^{-2}\big(\mathrm{Ai}(x+y)-\mathrm{Ai}(x-y)\big)^2\,.
\end{equation}
We split the $y$-integration into the intervals $[0,c]$, $ [c,\infty)$.  In $[0,c]$ we Taylor expand 
$\mathrm{Ai}(x+y)-\mathrm{Ai}(x-y)$ in $y$ and choose $c$ so small that the first order dominates.
We then determine $c_1$ such that $(\mathrm{e}^y-1)^{-2} \leq c_1 \mathrm{e}^{-y}$ on $ [c,\infty)$.
Using Schwarz inequality yields the estimate
\begin{eqnarray}\label{D.5}
 &&\hspace{10pt}\mathrm{tr}|A_2|^2\leq c_2 \int^\infty_s \mathrm{d}x\Big(\int^c_0
\mathrm{d}y y^2 (\mathrm{e}^y-1)^{-2} \mathrm{Ai}'(x)^2\nonumber\\
&&\hspace{22pt}+2 \int^\infty_0 \mathrm{d}y\mathrm{Ai}(x+y)^2+
2\int^\infty_0 \mathrm{d}y \mathrm{e}^{-y}\mathrm{Ai}(x-y)^2\Big)
\end{eqnarray}
for a suitable choice of the constant $c_2$. By definition
\begin{equation}\label{D.6}
\mathrm{tr} |A_3|^2=\int^\infty_s \mathrm{d}x \int^\infty_0
\mathrm{d}y \mathrm{e}^{-y} \mathrm{Ai}(x-y)^2
\end{equation}
and, repeating the argument for $\mathrm{tr} |A_2|^2$,
\begin{eqnarray}\label{D.7}
&&\hspace{-32pt}\mathrm{tr} |A_4|^2=\int^\infty_s \mathrm{d}x
\int^\infty_0 \mathrm{d}y (\mathrm{e}^{y}-1)^{-2} \mathrm{e}^y
\big(\mathrm{Ai}(x+y)-\mathrm{Ai}(x-y)\big)^2\nonumber\\
&&\hspace{10pt} \leq c_2\int^\infty_s \mathrm{d}x \Big(\int^c_0
\mathrm{d}y y^2 (\mathrm{e}^y-1)^{-1} \mathrm{e}^y \mathrm{Ai}'(x)^2\nonumber\\
&&\hspace{22pt}+2 \int^\infty_0 \mathrm{d}y\mathrm{Ai}(x+y)^2+
2\int^\infty_0 \mathrm{d}y \mathrm{e}^{-y}\mathrm{Ai}(x-y)^2\Big)\,.
\end{eqnarray}

Using that $|\mathrm{Ai}(x)|\simeq\exp[-x^{3/2}]$, $|\mathrm{Ai}'(x)|\simeq\exp[-x^{3/2}]$
for $x\to\infty$ and $|\mathrm{Ai}(x)| \simeq
|x|^{-1/4}$, $|\mathrm{Ai}'(x)|\simeq
|x|^{1/4}$ for $x\to-\infty$, all integrals are bounded with the
asymptotics $\exp[-s^{3/2}]$ for $s\to\infty$ and $|s|^{3/2}$ for
$s\to-\infty$.

Let us consider the integral (\ref{4.27}) for $\rho_t(s)$. The first factor decays as $\mathrm{e}^{-u}$ for $u\to\infty$ and as
$\exp[-\mathrm{e}^{|u|}]$ for $u\to-\infty$. For the second factor we use the inequality
$|\det(1+B)|\leq
\exp[\mathrm{tr}|B|]$, see \cite{RS}, Section XIII:17. From our previous estimates on the trace norm of $B_t$,
$g_t(u)$ is bounded by $c$ for $u\to\infty$ and as
$c\exp[|u|^{3/2}]$ as $u\to-\infty$, which establishes
integrability. For the integral  (\ref{4.25}) defining $F_t(s)$ one
uses that for large $u$ the determinants behave as
$1+\mathcal{O}(\exp[-u^{3/2}])$.
\end{proof}

\end{appendix}

\end{document}